\documentstyle[12pt]{article}

  \newcommand*\ti[5]{{\em #5}, {#1} {\bf #2}, #3 (#4)}

\newcommand*\jhep{JHEP}
\newcommand*\np{Nucl. Phys.}
\newcommand*\pl{Phys. Lett.}

\if@twoside  m
\else
\fi
 1
 1
\font\sr = msbm10 scaled \magstep 1

\def\BR{\mbox{\sr R}}

\def\LL{{\cal L}}

\def\DD{{\cal D}}

\def\id{\mbox{id}}

\def\beq{\begin{equation}}
\def\eeq{\end{equation}}
\def\beqa{\begin{eqnarray}}
\def\eeqa{\end{eqnarray}}

\begin{document}
\begin{titlepage}
\rightline{MPI}
\rightline{LMU-TPW 2000-04}
\rightline{hep-th/0001032v3}
\vspace{4em}
\begin{center}

 {\LARGE{\bf Noncommutative Yang-Mills from equivalence of star products}}

 \vskip 1.5cm

 {{\bf Branislav Jur\v co${}^{*}$ and Peter Schupp${}^{**}$ }}

 \vskip 0.5 cm

${}^*$Max-Planck-Instiut f\"ur Mathematik\\Vivatgasse 7\\
D-53111 Bonn, Germany\\[1ex]
${}^{**}$Sektion Physik\\
Universit\"at M\"unchen\\
Theresienstr.\ 37\\
D-80333 M\"unchen, Germany

 \end{center}

 \vspace{1 cm}

 \begin{abstract} It is shown that the transformation between ordinary and 
noncommutative Yang-Mills theory as formulated by Seiberg and Witten
is due to the equivalence of certain star products on the D-brane world-volume.
 \end{abstract}

\vfill
\noindent 
January 2000  \vspace{4pt}
\hrule \vspace{4pt}
\hbox{{\small{\it e-mail: }}{\small\quad jurco@mpim-bonn.mpg.de 
 ,\quad schupp@theorie.physik.uni-muenchen.de}}
\end{titlepage}\vskip.2cm

\newpage

\setcounter{page}{1}
\section{Introduction}
The noncommutativity of coordinates in D-brane physics has lately received considerable 
attention. See \cite{SW} and references therein, in particular \cite{CDS,CH,DH,MZ,W}. 
It was examined thoroughly from different points of view. 
On one side the transverse
coordinates of $N$ coinciding D-branes are described by $N\times N$ matrices on the other
side the end points of an open string become noncommutative in the presence of a constant 
B-field.   
We shall not go into details here and just mention a fact that is most relevant to the 
present letter: In both situations D-branes in the presence of a large background gauge 
field can be equivalently described by either
commutative or noncommutative gauge fields.

In this letter we will consider the problem from the D-brane world-volume 
perspective. 
The idea is the following: We formulate the problem within 
the framework of symplectic
geometry and Kontsevich's deformation quantization to obtain abstract but general results
independent of particularities of specific (path integral) quantizations 
\cite{Ishibashi,Okuyama,AD}. 
An equivalence of certain star products
will lead us to a transformation between two quantities, which physically 
can be interpreted as ordinary and non-commutative Yang-Mills fields. 
Within this approach
an existence of such a relation is a priori guaranteed. 
We then show that such a transformation is
necessarily identical to the transformation proposed by Seiberg and Witten \cite{SW}. 
All this can be done rigorously.
In the last part of the letter we will discuss how all this is related 
to the formulation that uses a path integral 
representation of boundary states \cite{Ishibashi, Okuyama}.

\section{Classical description}
\label{classical}

For the classical description of the problem the following lemma of 
Moser~\cite{Moser} is crutial. 
Let $M$ be a symplectic manifold and  
$\omega=\omega_{ij}(x)\mbox{d}x^i\wedge \mbox{d}x^j$ the 
symplectic form 
on $M$. The symplectic form is closed $\mbox{d}\omega=0$ and its coefficient 
matrix nondegenerate $\mbox{det}\, \omega_{ij}(x)\neq 0$ for all $x\in M$.
If $\omega'$ is another symplectic form on $M$ such that it 
belongs to the same cohomology class
as $\omega$ and if the $t$-dependent form ($t \in [0,1]$)
\beq
\Omega=\omega + t(\omega'-\omega)
\eeq
is nondegenerate, then
\beq
\omega'-\omega=\mbox{d}a
\eeq
for some 1-form $a$,
the $t$-dependent vector field $X$, implicitly given by
\beq
i_{X}\Omega+a=0
\eeq
is well defined and
\beq
\LL_{X}\Omega \equiv 
\mbox{d}(i_{X}\Omega)+i_{X}\mbox{d}\Omega+\partial_t\Omega=-\mbox{d}a +
(\omega'-\omega)=0.
\eeq
This implies that all $\Omega(t)$ are related by coordinate transformations generated
by the flow of $X$: $\rho^*_{tt'}\Omega(t')=\Omega(t)$, where
$\rho^*_{tt'}$ is the flow of $X$. Setting
$\rho^*=\rho^*_{01}$ we have in particular
\beq
\rho^*\omega' = \omega.
\eeq
Explicitely
\beq
\rho^* = \left.e^{\partial_t + X}e^{-\partial_t}\right|_{t=0} 
= e^{\theta^{ij}a_j\partial_i - \frac{1}{2}\theta^{ik}f_{kl}\theta^{lj} a_j
\partial_i + o(\theta^3)},
\eeq
where  $\theta^{ij} \omega_{jk} = \delta^i_k$ and
$f_{kl} = \partial_k a_l - \partial_l a_k$.

The only complication is that $X$ may not be complete, which is no problem for $M$ compact.
For noncompact $M$ (in our case an open domain in $\BR^{2n}$) we have to treat $t$ as
a formal parameter and work with formal diffeomorphisms given by formal power series in $t$.
Specifying
$t=1$ amounts to considering formal power series in the matrix elements of 
$(\omega'-\omega)$. This is the same as 
assuming that $\mbox{d}a$ is small or $\omega$ large. 
Alternatively we could work with formal power series in $\theta^{ij}=\omega^{-1}_{ij}$. 
In either case $\Omega$ is nondegenerate. In this sense
we always have a coordinate change on $M$ which relates the two symplectic 
forms $\omega$ and $\omega'$. In the cases $t=0$ and $t=1$ we denote the 
Poisson brackets
by $\{ , \}$ and $\{ , \}'$ respectively.

Consider now a gauge transformation 
$a\mapsto a +\mbox{d}\lambda.$
The effect upon $X$ will be
\beq
X\mapsto X+X_{\lambda},
\eeq
where $X_{\lambda}$ is the Hamiltonian vector field
\beq
i_{X_{\lambda}}\Omega+\mbox{d}\lambda=0
\eeq
and $\LL_{X_{\lambda}} \Omega = 0$.
 The whole transformation induced by $\lambda$,
including the coordinate transformation $\rho^*$ corresponding to $a$ is
\beq
f \stackrel{(\lambda)}{\mapsto} f + \{\tilde\lambda,f\}' \stackrel{(a)}{\mapsto}
\rho^* f + \{\rho^*\tilde\lambda,\rho^* f\},
\eeq 
where we have used $\rho^*\{\tilde\lambda,f\}' = \{\rho^*\tilde\lambda,\rho^*f\}$.
We shall give an expression of $\tilde\lambda$ in the case of constant
$\theta$ in section~4.

Physically we can view the above coordinate transformations
either as active or passive, i.e.\ we either have two 
different symplectic structures $\omega$, $\omega'$
on the same manifold related by an active 
transformation or we have just one symplectic structure expressed 
in different coordinates. The additional infinitesimal canonical
transformation does not change the symplectic structure.
Let us mention that the paper \cite{Cornalba} is in fact an 
explicit realization of the Moser lemma in the 
situation describing a D-brane in the background gauge field.

\section{Deformation quantization}

We would now like to consider the deformation quantization \cite{BFFLS} of the
two symplectic structures $\omega$ and $\omega'$ a la Kontsevich. 
We follow the definitions
and conventions of~\cite{Kontsevich}. 

The set of equivalence classes of Poisson structures on a smooth 
manifold $M$ depending formally on $\hbar$,
\beq
\alpha(\hbar)=\alpha_1\hbar +\alpha_2\hbar^2+ \ldots, \qquad
[\alpha,\alpha]=0,
\eeq
where $[,]$ is the Schouten-Nijenhuis bracket of 
polyvector vector fields, is defined modulo the action of 
the group of formal paths in the diffeomorphism group of $M$, 
starting at the identity diffeomorphism.
Within the framework of Konstevich's deformation quantization the 
equivalence classes of Poisson manifolds can be 
naturally identified with the sets of gauge equivalence 
classes of star products on the smooth manifold~$M$.
The Poisson structures $\alpha$, $\alpha'$ can be identified with the
series $\alpha(\hbar) = \hbar \alpha$ and $\alpha'(\hbar) = \hbar \alpha'$,
and via Kontsevich's construction with canonical gauge equivalence 
classes of star products. In view of Moser's Lemma
the resulting star products will also be equivalent in the sense 
of deformation theory. 

Since the two star products $*$ and $*'$ on $M$, corresponding
to $\alpha(\hbar)$ and $\alpha'(\hbar)$, are equivalent,
there exists an automorphism 
$D(\hbar)$ of $A[[\hbar]]$, which is a formal 
power series in $\hbar$, starting with
the identity, with coefficients that are 
differential operators on $A \equiv C^{\infty}(M)$,
such that for any two smooth functions $f$ and $g$ on $M$ 
\beq
f(\hbar)*' g(\hbar)=D(\hbar)^{-1}(D(\hbar)f(\hbar)*D(\hbar)g(\hbar)).
\label{equi}
\eeq
Note, that we first have to take care of the classical part of the
transformation via pullback by $\rho^*$, so that the remaining
automorphism $D(\hbar)$ is indeed the identity to zeroth order in $\hbar$.
The complete map, including the coordinate transformation is
$\DD = D(\hbar) \circ \rho^*$.

The inner automorphisms of $A[[\hbar]]$, given by similarity transformation
\beq
 f(\hbar) \mapsto \Lambda(\hbar)*f(\hbar)*(\Lambda(\hbar))^{-1} , \label{sym}
\eeq
with invertible $\Lambda(\hbar)\in A[[\hbar]]$, do not change the star product. 
Infinitesimal transformations
that leave the star product invariant are
necessarily derivations of the star-product.
The additional gauge transformation freedom $A \rightarrow A + d\lambda$ in
Moser's lemma induces an infinitesimal canonical transformation and,
after quantization, an inner derivation of the star product $*'$. 
We will use the fact that this transformation (including classical and quantum
part) can be chosen as
\beq
f \mapsto f+ i \tilde\lambda *' f - i f *' \tilde\lambda.   \label{qmap}
\eeq
This, as we shall see, directly lead
to the celebrated relation of a noncommutative gauge transformation.

We shall not try to review deformation quantization and Kontsevich's formula for the 
star product in its full generality here.
A detailed description is given in the original paper \cite{Kontsevich}, the path integral
representation using a topological sigma-model on the disc was developed 
in \cite{CattaneoFelder} and an excellent historical overview of deformation quantization
and many references can be found in \cite{Sternheimer}. 
We only note that in the case of constant 
Poisson tensor $\alpha$ one obtains the well-known Moyal
bracket. We are, however, interested in the existence of a natural 
star product for any Poisson manifold, which is guaranteed according to Kontsevitch.
In this letter we technically only use the corresponding result for symplectic manifolds 
\cite{Dewilde,Fedosov,OMY}.

In the following we will absorb $\hbar$ in $\theta$, $\theta'$, etc.

\section{Seiberg-Witten transformation}

To make contact with the discussion of Seiberg and 
Witten we take $\omega$ to be the symplectic form
on $\BR^{2n}$, the D-brane world-volume, induced by a constant $B$-field:
\beq
\omega=\theta^{-1}_{ij}\mbox{d}x^i\wedge\mbox{d}x^j
\eeq
with 
\beq
\theta^{ij}=\left(\frac{1}{g+B}\right)^{ij}_A; \label{0}
\eeq
$g$ is the constant closed string metric and the subscript A 
refers to the antisymmetric part of a matrix
(we have set $2\pi\alpha'=1$).
In the zero slope limit 
\beq
\omega=B.
\eeq
For $\omega'$ we take
\beq
\omega'\equiv (\theta')^{-1}_{ij}\mbox{d}x^i\wedge\mbox{d}x^j=\omega + F,\label{0'}
\eeq
where $F=F_{ij}\mbox{d}x^i\wedge \mbox{d}x^j$ is the field strength of the
rank one gauge field $A$. (The
extension of the following to higher rank is straightforward.)
We are in the situation of section~\ref{classical}, with $a=A$ 
being the gauge field.
The star products  induced by 
Poisson structures $\theta$ and $\theta'$
are equivalent and the equivalence transformation (including the classical part)
is given by the map $\DD = D \circ \rho^*$, where $\rho^* = \id + 
\theta^{ij} A_j \partial_i
+ \frac{1}{2} \theta^{kl} A_l \partial_k \theta^{ij} A_j \partial_i
+\frac{1}{2} \theta^{kl} \theta^{ij} A_l F_{kj} \partial_i + o(\theta^3)$, 
see also \cite{Cornalba};
$D$ acts trivially on $x^i$ to this order, but can of course in principle be computed 
order by order to any order in $\theta$.
It is convenient to write the result of $\DD$ acting on the 
coordinate functions $x^i$
in the form \cite{Cornalba, Ishibashi, Okuyama}
\beq
\DD x^i =x^i + \theta^{ij}\hat{A}_j
= x^i + \theta^{ij} a_j +
\frac{1}{2} \theta^{kl}   \theta^{ij} A_l (\partial_k A_j)
+\frac{1}{2} \theta^{kl} \theta^{ij} A_l F_{kj}  + o(\theta^3)\label{xtr}
\eeq
with $\hat A$ a function of $x$ depending on $\theta$, $A$ and derivatives of $A$, as 
shown.   
It is obvious that $\hat A$ has the form 
$\hat A = A+ o(\theta) + \ldots$, since to lowest order in $\theta$ it has to 
reproduce the coordinate transformation $\rho^*$ relating the two symplectic forms
$\omega$ and $\omega'$.

Let us now discuss what effect a gauge transformation $A\mapsto A +d\lambda$ has
in this picture:
It represents the freedom in the choice of symplectic potential
$A'=\frac{1}{2}\omega_{ji}x^j\mbox{d}x^i + A$ for $\omega'.$
In section~\ref{classical} we found that classically the 
gauge transformation amounts to an infinitesimal 
canonical transformation, and, after deformation 
quantization, it has the form (\ref{qmap}).
The whole map is
\beq
f \stackrel{(\lambda)}{\mapsto} f+ i \tilde\lambda *' f - i f *' \tilde\lambda
\stackrel{(A)}{\mapsto} \DD f+  i \DD \tilde{\lambda} *\DD f 
-i \DD f*\DD \tilde{\lambda} .\label{df}
\eeq
Let us introduce $\hat{\lambda}$ as a shorthand for $\DD \tilde{\lambda}$.  
$\hat{\lambda}$ obviously depends on $\theta$, $A$, 
derivatives of $A$ and the classical gauge transformation $\lambda$.
Explicitely:
\beq
\tilde \lambda = \lambda - \frac{1}{2} \theta^{ij} A_j (\partial_i \lambda) + o(\theta^2), 
\qquad \hat \lambda =  \lambda + \frac{1}{2} \theta^{ij} A_j (\partial_i \lambda)
+ o(\theta^2).
\eeq
We would like to express the result of the map 
(\ref{df}) acting on the coordinates $x^i$ again in the form (\ref{xtr}), but
with $\hat A_j$ replaced with $\hat A_j + \delta\hat A_j$. Using (\ref{xtr}) and
$x^i*x^j - x^j*x^i = i\theta^{ij}$ to compute the $*$-commutator 
$[\hat \lambda \stackrel{*}{,} x^i]$, we find
\beq
\delta \hat{A}_i 
= \partial_i \hat{\lambda} + i\hat{\lambda}*\hat{A}_i -i\hat{A}_i*
\hat{\lambda}\label{gauge}.
\eeq
We see, as expected, that the relation between $\hat{A}$ and $A$ implied by 
the coordinate transformation
(\ref{xtr}) is precisely the 
same as the one proposed by Seiberg and Witten based on the expectation 
that an ordinary gauge transformation on $A$ 
should induce a noncommutative gauge transformation (\ref{gauge}) on $\hat{A}$.
We furthermore see that within the framework of deformation quantization a la Kontsevich 
the existence of such a transformation 
between the commutative and noncommutative descriptions is
guaranteed. It is not hard to compute the terms of higher order in $\theta$
directly in our approach.

In essence the Seiberg-Witten transformation between 
the commutative and noncommuative description of D-branes
is possible due to equivalence of two star products, namely the one defined by the Poisson 
tensor $\theta$ (\ref{0}) and the another one defined by the Poisson $\theta'$ (\ref{0'}).

Let us remark that we make contact here with another approach to
noncommutative gauge theory \cite{Wess}, whose relations and manipulations
resemble the ones of 
this section, but with a different philosophy.  Equation (\ref{xtr}) 
defines a covariant coordinate in that theory.

\section{Relation to boundary states formalism}

The string boundary state coupled to the $U(1)$ gauge 
field admits a path 
integral representation. Let $|D\rangle$ be a Dirichlet boundary state, $X^i(\sigma)|D\rangle=0$ at some 
fixed instant $\tau=0$. $X^i$ are the string coordinates and $P_i$ the conjugate momenta.
The boundary state $|B\rangle$ coupled to a U(1) gauge field $A'$ is then given as
\beq
|B\rangle =\int \DD x \,\mbox{exp}({i\int \mbox{d}\sigma
A_i'(x)\partial_{\sigma}x^i}-P_ix^i)|D\rangle .
\eeq
The path integral itself can be interpreted within the framework of Kontsevich 
deformation quantization \cite{CattaneoFelder}: If we denote $*'$ the star product obtained a la Kontsevich
from $\omega'=\mbox{d}A'$ then the path integral is the trace of the path-ordered exponential
$\mbox{P}[\exp(-i\int \mbox{d}\sigma P_ix^i)]_{*'}$, where we assume implicitly 
the $*'$-product within the 
exponential. In the notation of the previous section
$A'=\frac{1}{2}B_{ij}x^i\partial_\sigma x^j + A$. Let us translate the gauge equivalence
of the star products $*'$ and $*$ to the language of boundary states. We get
the condition
\beqa
\lefteqn{\int \DD x \,\exp\left(i\int \mbox{d}\sigma 
\frac{1}{2}B_{ij}x^i\partial_\sigma x^j + 
A -P_ix^i\right)|D\rangle} && \nonumber \\
&& = \int \DD x \,\exp\left(i\int \mbox{d}\sigma \frac{1}{2}B_{ij}x^i
\partial_\sigma x^j  
-P_i(x^i+\theta^{ij}\hat A_j)\right)|D\rangle.
\eeqa
This is exactly the condition of \cite{Okuyama}. The above equality is evidently true even
without path integrals on its both sides acting on the Dirichlet boundary state.

\section*{Acknowledgements}

We would like to thank S.\ Theisen for many helpful discussions. B.J.\ thanks
the Alexander-von-Huboldt-Stiftung for support and P.\ Xu and P.\ \v Severa 
for discussions.


\begin{thebibliography}{article}

\bibitem{SW} N. Seiberg, E. Witten, \ti{\jhep}{9909}{032}{1999}{String Theory
and Noncommutative Geometry}, hep-th/9908142.

\bibitem{CDS} A. Connes, M. R. Douglas, A. Schwarz,
\ti{\jhep}{9802}{003}{1998}{Noncommutative Geometry and Matrix Theory:
Compactification on Tori}, hep-th/9711162.

\bibitem{CH} C.-S. Chu, P.-M. Ho, \ti{\np}{B550}{151}{1999}{Noncommutative open
string and D-brane}, hep-th/9812219.

\bibitem{DH} M. R. Douglas, C. Hull, \ti{\jhep}{9802}{008}{1998}{D-Branes and the
noncommutative torus}, hep-th/9711165.

\bibitem{MZ} B. Morariu, B. Zumino, in: Relativity, 
Particle Physics and Cosmology, World Scientific, Singapore, 1998, hep-th/9807198.

\bibitem{W} W. Taylor, 
\ti{\pl}{B394}{283}{1997}{D-Brane field theory on compact spaces}, hep-th/9611042.

\bibitem{Ishibashi} N. Ishibashi, \emph{A Relation between Commutative 
and Noncommutative Descriptions of D-branes}, hep-th/9909176.

\bibitem{Okuyama} K. Okuyama, \emph{A Path Integral Representation of the 
Map between Commutative and Noncommutative Gauge Fields}, hep-th/9910138.

\bibitem{AD} O. Andreev, H. Dorn, {\em On open string sigma model and
noncommutative gauge fields}, hep-th/9912070.

\bibitem{Moser} J. Moser, \ti{Trans. Amer. Math. Soc.}{120}{286}{1965}{On the volume 
elements on a manifold}.

\bibitem{Cornalba} L. Cornalba, \emph{D-brane Physics and 
Noncommutative Yang-Mills Theory}, hep-th/9909081.

\bibitem{BFFLS} F. Bayen, M. Flato, C. Fronsdal, A. Lichnerowicz, D.
Sternheimer, \ti{Ann. Physics}{111}{61}{1978}{Deformation theory and
quantization. I. Deformations of symplectic structures}.

\bibitem{Kontsevich} M. Kontsevitch, \emph{Deformation quantization of Poisson
manifolds, I,}\\q-alg/9709040; \ti{Lett. Math. Phys.}{48}{35}{1999}{Operads 
and Motives in Deformation Quantization}, math/9904055.

\bibitem{CattaneoFelder} A. S. Cattaneo, G. Felder, \emph{A path integral 
approach to the Kontsevich quantization formula}, math/9902090.

\bibitem{Sternheimer} D. Sternheimer, \emph{Deformation Quantization: Twenty
Years After}, \\math/9809056.


\bibitem{Dewilde} M. De Wilde, P.B.A. Lecomte, \ti{Lett. Math.
Phys.}{7}{487}{1983}{Existence of star-products and of formal deformations of a
Poisson Lie algebra of arbitrary symplectic manifolds}.

\bibitem{Fedosov} B. Fedosov, \ti{J. Diff. Geom.}{40}{213}{1994}{A simple
geometric construction of deformation quantization}.

\bibitem{OMY} H. Omori, Y. Macda, A. Yoshioka, \ti{Advances in Math.
(China)}{85}{224}{1991}{Weyl manifolds and deformation quantization}.

\bibitem{Wess} J. Madore, S. Schraml, P. Schupp, J. Wess,
{\em Gauge Theory on Noncommutative Spaces}, hep-th/0001203.

\end{thebibliography}
\end{document}